\documentclass[a4paper,10pt]{article}
\usepackage{graphicx}

\begin{document}
\pagestyle{plain}
\setcounter{page}{0}

\parindent 8mm
\pagestyle{empty}

\vspace*{2cm}
\title{Node Exchange Network and its Statistical Analysis}
\author{ Norihito Toyota}
\date{Hokkaido Information University,
59-2 Nishinopporo Ebetsu City, Japan \\
E-mail:toyota@do-johodai.ac.jp
}

\maketitle
\baselineskip 5mm
\pagestyle{plain}
\begin{abstract}
In considering a social network, there are cases where people is transferred to another place. Then the physical (direct) relations among nodes are lost by the movement. In terms of a network theory, some nodes break the present connections with neighboring nodes, move and there build new connections of nodes.  For simplicity we here consider only that two nodes exchange the place  each other on a network. Such exchange is assumed to be constantly carried out.  We study this dynamic network (node exchange network NEN) and uncover some new features which usual networks do not contain. We mainly consider average path length and the diameter. Lastly we consider a propagation of one virus on the network by a computer simulation.  
They are compared to other networks investigated hitherto. The relation to a scale free network is also discussed.

\end{abstract}
\bf{key words:} 
\it{Node Exchange Network, Smallworld Network, Scale Free Network, Average path  length. }

\rm
\normalsize
\baselineskip 6mm
\pagestyle{plain}

\renewcommand{\thefootnote}{\fnsymbol{footnote}} 
\rm
\normalsize
\baselineskip 6mm

\section{Introduction}

In a social network, we need to consider the possibility that people is transferred to another place. 
Then the physical (direct) relations among them are lost by the movement. 
In terms of a network theory, this means that some nodes break the present connections with neighboring nodes, move and there build new connections of nodes.  For simplicity we here consider only that two nodes exchange the place each other on a network. Such exchange is assumed to be constantly carried out.  We study this dynamic network (node exchange network NEN) to uncover some new features which usual networks do not contain. 
Then we introduce a handy network that has essentially same properties as random networks, instead of random networks.   
We mainly consider average distance between any pairs of nodes and  the diameter. 
Lastly we study the phenomenology by computer simulation that one virus on the network spreads throughout  the network.  
They are compared to other networks including the handy network, regular lattice and the small world network (SW-NET)\cite{Watt1},\cite{Watt2}, and furthermore the relation to a scale free network (SF-NET)\cite{Albe1},\cite{Albe2},\cite{Albe3}, is discussed.

\section{Node Exchange Network; NEN} 
As explained in the previous section, we consider that nodes exchange each other on a regular network. 
This network seems to look like a small world network. 
However it is necessarily not the case. 
By the movement, 
the nodes and the edges accompanied with them are entirely cut, and the nodes are connected with new edges each other.  
Notice that a network topology is apparently invariant under the procedure. 
In  small world networks the  static properties are only pursured but  
the  dynamic properties are rather important in the NEN.   

The alghorism for formulating this network is as follows;\\
1. Prepare a regular (one dimensional )network with a periodic boundary condition. This is like a ring. \\
2. Randomly choose two nodes on it at random and exchange them. 
This procedure is repeated $Q$ times.\\
3. Evaluate correct quantities of the network. \\
4. 1$\sim$3, which is one round,  is repeated $M$ times.   \\  
  
In such a way the network is analyzed dynamically.

\section{Results of Simulation }
We show the results of computer simulations. 
First of all we evaluate a diameter $D$ and the average distance between any pairs of nodes $L$ of the NEN . 
Both of them should show the same behaviours essentially in static networks. 
In the case, however, it may necessarily not be so, because to measure $D$ take more  steps than to measure  $L$ and the networka are essentially differnt after different time steps  in a dynamic network.  

Before doing computer simulation, we conveniently introduce a network "random graph with fixed degree", RNFD. 
Usualy the degree distribution of random networks is the Poisson distribution in the limit of a large network size $n \rightarrow \infty$. 
Then a diameter behaves as $\frac{\log n}{\log <k>}$ where $<k>$ is an average degree\cite{Chun}.    
This behaviour, however,  only appears in the large $n$ and so we introduce more handy model with the same properties essentially as random networks. 
This is  RNFD that exhibits essentially the same behaviours as random networks for smaller $n$. 
This is shown by doing simulation later.  
In RNFD, each node has exactly same number of links, that is the degree $k$, but they are randomly connected with other nodes. 
In terms of an adjacency matrix, it is a symmetric matrix with randomly $k$ elements of 1 in each column and row.  
We use this handy model to study network properties after this.

 Fig.1 shows diameters $D$ vs. size $n$ of the NEN constructed from degree $k=4$ regular lattice  and  RNFD with degree $k=4$, respectively. 
In Fig.1, points and curved lines show simulation data and approximate curve, respectively.    
This show that $n$ dependence of $D$ in the NEN is exponential, while that in RAFD is logarithmic such as random networks. 
(Notice that the coeffient of $\log$  is not $1/\log _e k$ such as random graphs. 
This result also means that RNFD is not so small world compared to random networks. )    
Their properties are invariant under changing $Q$ in NEN or $k$ in RNFD.  
Since it is linear  $D=\frac{n}{2k}$ in one dimensional regural lattice with a periodic boundary condition, NEN  is a network intermediate between a regular lattice and random networks such as SW-NET.  


\addtocounter{figure}{0}
\begin{figure}[b]
\begin{center}
\includegraphics[scale=1.0]{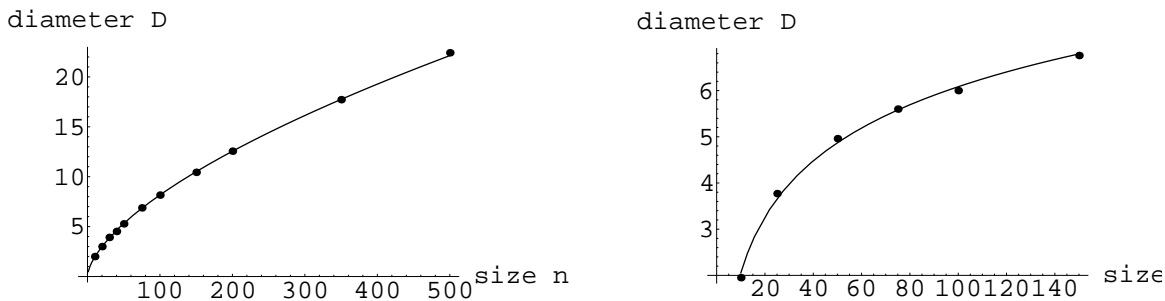}
\caption{Diameters of the NEN with $Q=10$ for average of $50$ times (left) and RNFD with $k= 4$ for average of $100$ times (right). Approximate formula of them are $D=0.4725 n^{0.619}$ and $D=1.7507 \log _e n -1.9778$, respectively. }
\end{center}
\end{figure}

To  clear the point we study average distance between any pairs of nodes, which is well investigated in SW-NET and SF-NET. 
Fig.2 shows $L-n$ curves of RNFD with $k=4$ and NEN with $Q=10$ and $k=4$.  
Essentially $L$ has same property as $D$. 
The reason will be that the number of steps needed up to a complete estimation of $D$ is nearly equal to that of $L$.  
This properties is invariant under changing $Q$ in the NEN or $k$ in a random network. 
A regural lattice shows linear dependence in $L-n$ relation such as $D-n$. 
Then  NEN is not so small world and so large world. 
Fig. 3 shows theoretical $D-n$ curves of SW-NET\cite{Newm1} and SF-NET\cite{Boll},\cite{Cohe} that are given by 
\begin{equation}
L(n)=\left\{
\begin{array}{ll}
\frac{\log(4np)}{8p} & \mbox{ for $2np >> 1$ and SW-NET}, \\
\frac{\log_e n}{ \log_e \log_e n} & \mbox{ for SF-NET},
\end{array}
\right.
\end{equation}
and their approximation curves. 
The approximation curves, which is useful for comparing to NEN or random network phenomenologically, fit too well to distinguish theoretical curves and approximation ones  up to $n=1500$. 
$p$ is the renwiring possibility of links in SW-NET and $p=0.05$ is taken in Fig.3.   
In SF-NET, both of exponential and logarithmic functions fit almost perfectly. 
Though it is possible that both of NEN and SF-NET can be approximated by exponential functions, they are  very different from each other in index.   
Thus NEN does not show so small world property. This property is invariant under changing $Q$ value. 
As we increse $Q=1,\;5,\;  10, ...$, the index decreses to $s=\;0.83,\; 0.62,\;0.58,...$. 
$s=0.07$ in SF-NET is different from those of NEN in order 
(we should interpret that the small $s$ means logarithmic function) 
and it  seems not to be able to overcome  the difference. 
Thus SF-NET and NEN are essentially thought to be different networks. 

\addtocounter{figure}{0}
\begin{figure}[t]
\begin{center}
\includegraphics[scale=1.0]{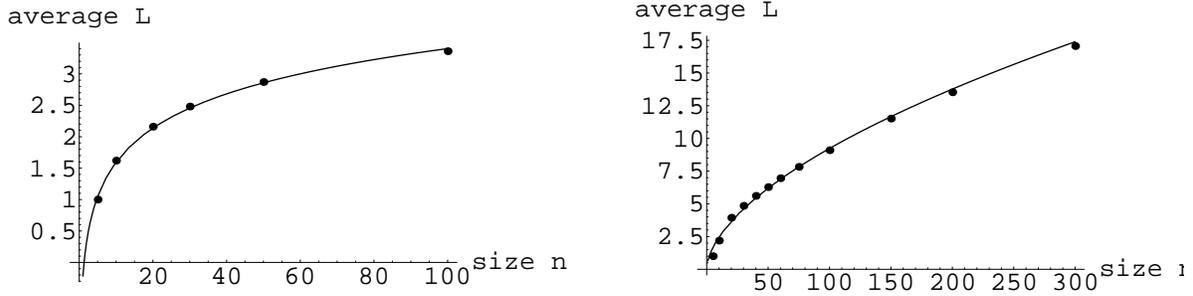}
\caption{Average distances between two nodes for average of $100$ times: The left is an average $L$ of RNFD with $k=4$. The bottom is that of NEN with $Q=10$. The approximate formula of them are $L=0.7861 \times \log_e n- 0.2182$ and $L= 0.6509 \times n^{0.579}$, respectively}
\end{center}
\end{figure}

\addtocounter{figure}{0}
\begin{figure}[t]
\begin{center}
\includegraphics[scale=1.0]{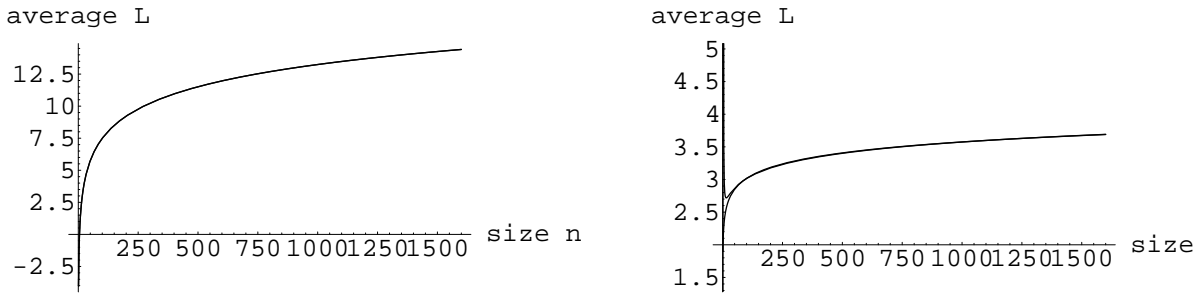}
\caption{Average distances between two nodes: The left is an average $L$ of SW-NET with $k=4$ and $p=0.05$, and the right is  that of SF-NET.  The approximate formula of them are $L=2.5 \log_e n- 4.0236$ and $L= 0.2421 \log_e n + 1.9031$ or $L=2.1888 \times n^{0.0707}$, respectively}
\end{center}
\end{figure}

The clustering coeffient and the degree distribution  have no significance in dynamic networks, since the network topology in NEN is apparently invariant temporally so that it takes the same value as the one of regular lattice. As for this we may have to introduce a sort of new kind of index to investigate NEN in more details. 

We next study a propagation on this network, NEN. 
We  set a virus on a node chosen at random. 
The virus propagates on the network so that all nodes will get infected with the virus after some time steps $S$. 
Thus we study the $S$ that can more directly estimate connectibility of networks than $D$ and $L$.  
They are shown in Fig.4 where the propagation length is the number of steps untill all nodes are infected. 
This  corresponds to  cases that is  an infection rate $I =1$ and a cure rate $C=0$.    
A random network shows logarithmic behaviour but nearly linear in NEN. 
(We can not distinguish between two approximate formula up to our computer power.  )  
This also shows the crusial difference between a random network and NEN, they are are small world network in a sense.  
We notice that the results are rather unstable. 
It depends on whether the virus first put is layed on exchange node or not. 
Notice that we insist on studying only topological and ststistical properties of netoworks in this article, and more realistic studies of infections will be studied continuously.

\addtocounter{figure}{0}
\begin{figure}[h]
\begin{center}
\includegraphics[scale=1.0]{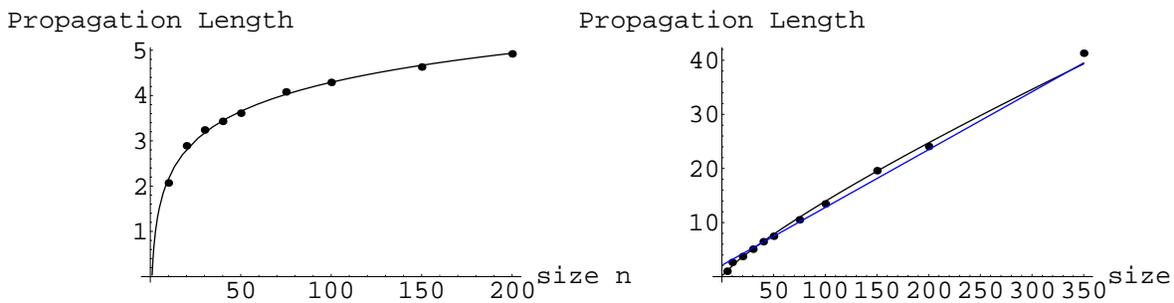}
\caption{Propagation Length $S$ for average of $150$ times: The left is $S$ of SW-NET with $k=6$, and the right is $S$ of NEN with $q=5$.  The approximate formula of them are $S=0.9253 \log_e n+ 0.0315$, and $S= 0.3069 n^{0.8285}$(black) or $S=0.1069  n + 2.1018 $(blue), respectively}
\end{center}
\end{figure}

\section{Weighted Node Exchange Net; WNEN}             
There are some similarities between the NEN and SF-NET apparently, we pursue this point still more.  
Scale free property appears from both of evolution and preferential attachment.  We introduce the idea  of the preferential attachment into  this dynamic NEN. 
We assume that the nodes which has been transferred once are transferred with high propability after that. 
At $m$ round and q times,  the probability  $p_i(t)$  that a node $i$ is chosen to be an exchange node at $t$ is assumed that
\begin{equation}
p_i(t)=\left\{
\begin{array}{ll}
\frac{1+p_i(t-1)N(t-1)}{N(t-1)+2}& \mbox{when the node $i$ is chosen as exchange node at time $t-1$ }, \\
p_i(t-1) & \mbox{others},
\end{array}
\right.
\end{equation}
where
\begin{equation}
N(t)=n+2t,\;\;\;\;t=m Q+q, \;\;\;\; 
p_i(0) =\frac{1}{N(0)}=\frac{1}{n} \; \mbox{ for all $i$}. 
\end{equation}

This reflects the fact that while active people often transfer, others trend to stay.        
We call this type of networks weighted Node Exchange Network (WNEN). 

The results of computer simulation of $L$, $D$ and a propagation of a virus on the network WNEN are just similar to those of the NEN. 
They are partly given in Fig.5.  
This is because the number of times of exchange till the evaluation of the above quantities  is completed is not so large. 
For that reason,  the  preferential attachment has not any drastic effects in the simulation. 

Thus we wish to prusue further characteristics of NEN or WNEN. 
We caluculate a sort of dynamic link number distribution $S(t)$.   
We, however,  need to notice that the topology of the NEN is invariant at each time step. 
The number of links connected with a node is same as the common degree of nodes every time. 
The $S_i(t)$ at times $t$ is the integration of nodes connected with the node $i$ in the past. The nodes that connected with the node $i$ twice or more in the past  is exempted from counting of $S_i(t)$. Thus $ S_i(t) \neq \sum_{t_0}^{t} k$. $S(t)$ is the average of all $S_i(t)$ for all nodes;
 
\begin{equation}
S(t)= \frac{\sum_{i=1}^n S_i(t) }{n}.
\end{equation}
We call this $S(t)$ an integral link number ILN.  
We explore the distributions for $S(t)$ and its time dependence. 
These time dependences are given in  Fig.6 where the horizontal axis means $S_i(m)$ and the vertical one means the logarithmic number of vertices with $S_i(m)$. Just at $m=10$, we can observe scale-free like behaviour. 
This is a reason that the exponential behaviours are observed in $L$ and so on at low $m$.   
As $m$ grows larger, it comes to break from the left to right, which show "ant-scale free" behaviour, in Fig.6. 
The term "anti-free" introduced above means that many nodes  enjoy a close intimacy with each other but a tiny minority   do not so, that is to say, dual to scal free. 
Thus following the WNEN temporally, we observe a series
$$ regular \rightarrow scale free \rightarrow random \rightarrow  anti-scale free \rightarrow regular.$$
Thus we observe a duality 
$$S_i(T-m) \leftrightarrow n-S_i(T+m),$$
where $T$ is nearly half time from initial time to the time that WNEN converges. 
\addtocounter{figure}{0}
\begin{figure}[t]
\begin{center}
\includegraphics[scale=1.0]{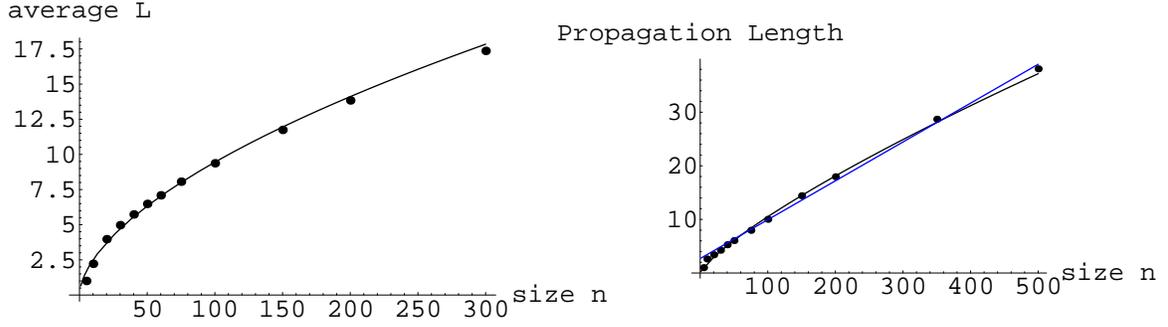}
\caption{$L$ for average of $100$ times and  $S$ for average of $150$ times in WENE with $q=10$ and $k=4$ : The left is $L$ that is approximately  $L= 0.6585 n^{0.5784}$ and the right is $S$ that is approximately $S=0.0725 n+2.6983$(black) or $S=0.2819 n^{0.7856}$(blue), respectively. }
\end{center}
\end{figure}

\addtocounter{figure}{0}
\begin{figure}[h]
\begin{center}
\includegraphics[scale=1.0]{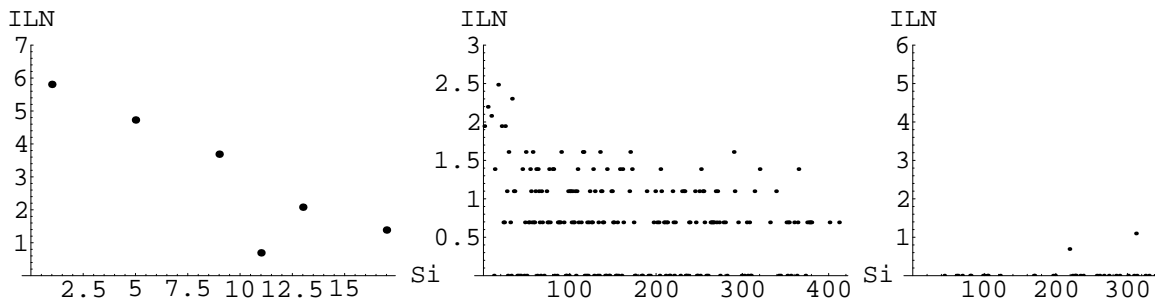}
\caption{Integral distributions of numbers of nodes on the WNE with $Q=10$ and $n=500$. The round numbers are $M=10,\; M=100,\; M=190$ from the left to the right, respectively.  }
\end{center}
\end{figure}

\section{Concluding Remarks}
By considering an effect that people often transfer in real society, we propose a network where nodes exchange each other dynamically. 
This network NEN where two nodes exhange the place constantly is a dynamic model in the sense that analyses are carried out during the node exchange. 
This is a different network from usual non-dynamic networks including various types of evolving networks where analyses are usually carried out after the enough growth.   
We evaluate diameters, averange path lengths between nodes and propagation lengths on networks including NEN by computer simulation. 
We explore furthermore WNEN that reflects preferential attachment in evolving network.  
As results of simulations it is proven that  NEN and WNEN have the middle properties between a random network and regular lattice model such as SW-NET.  
The essential prpperties, however, is rather  close to regular lattice model but clearly different from SW-NET.  
Though partially NEN also has the properties of scale free network, details should be more investigated from now on. 
The studies of properties of WNEN at large $t$ (middle or right cases in Fig.6) are also needed. 
Naturally  various real phenomena such as the propagation of pathogenic in more realistic setteng, robustness for fault, searching and so on  on NEN and WNEN 
should studied.



\end{document}